\documentclass[conference]{IEEEtran}
\IEEEoverridecommandlockouts
\usepackage{cite}
\usepackage{amsmath,amssymb,amsfonts}
\usepackage{algorithmic}
\usepackage{graphicx}
\usepackage{textcomp}
\usepackage{xcolor}
\def\BibTeX{{\rm B\kern-.05em{\sc i\kern-.025em b}\kern-.08em
    T\kern-.1667em\lower.7ex\hbox{E}\kern-.125emX}}
\begin{document}

\title{Hybrid Power-Law Models of Network Traffic
\thanks{This material is based upon work supported by the Assistant Secretary of Defense for Research and Engineering under Air Force Contract No. FA8702-15-D-0001, National Science Foundation CCF-1533644, and United States Air Force Research Laboratory Cooperative Agreement Number FA8750-19-2-1000. Any opinions, findings, conclusions or recommendations expressed in this material are those of the author(s) and do not necessarily reflect the views of the Assistant Secretary of Defense for Research and Engineering, the National Science Foundation, or the United States Air Force. The U.S. Government is authorized to reproduce and distribute reprints for Government purposes notwithstanding any copyright notation herein.}
}

\author{\IEEEauthorblockN{Pat Devlin$^1$, Jeremy Kepner$^2$, Ashley Luo$^2$, Erin Meger$^3$}
\\
\IEEEauthorblockA{$^1$Yale,  $^2$MIT, $^3$Universit\'e du Qu\'ebec \`a Montr\'eal}}

\maketitle

\begin{abstract}

The availability of large scale streaming network data has reinforced the ubiquity of power-law distributions in observations and enabled precision measurements of the distribution parameters.  The increased accuracy of these measurements allows new underlying generative network models to be explored.  The preferential attachment model is a natural starting point for these models.  This work adds additional model components to account for observed phenomena in the distributions.  In this model, preferential attachment is supplemented to provide a more accurate theoretical model of network traffic. Specifically, a probabilistic complex network model is proposed using preferential attachment as well as additional parameters to describe the newly observed prevalence of leaves and unattached nodes.  Example distributions from this model are generated by considering random sampling of the networks created by the model in such a way that replicates the current data collection methods.

\end{abstract}

\begin{IEEEkeywords}
networks, Zipf-Mandlebrot, preferential attachment, modelling
\end{IEEEkeywords}

\section{Introduction}

Recent events have underscored the increasing importance of the Internet to our civilization, necessitating a scientific understanding of this virtual universe \cite{hilbert2011world,li2013survey}. The pandemic induced a drastic increase in Internet usage due to activities such as remote education and work, streaming, social media, online shopping, and video games \cite{ali2020online, almaiah2020exploring, candela2020impact, pandey2020impact}. However, with this rise of online activity comes a potential surge of problematic internet usage that is made more urgent by the rising influence of adversarial Internet robots (botnets) on society \cite{allcott2017social,  BadBotReport, healey2018zero, ahmad2019botnets}.  Thus, for scientific, economic, and security reasons observing, analyzing, and modeling the Internet is essential.

  The two largest efforts to capture, curate, and share Internet packet traffic data for scientific analysis are  the Widely Integrated Distributed Environment (WIDE) project \cite{cho2000tr} and the Center for Applied Internet Data Analysis (CAIDA) \cite{claffy1999internet}.  These data have supported a variety of research projects resulting in hundreds of peer-reviewed publications \cite{CAIDApubs}, ranging from characterizing the global state of Internet traffic, to specific studies of the prevalence of peer-to-peer filesharing, to testing prototype software designed to stop the spread of Internet worms.  More recently, novel analysis of trillions of network observations enabled by high performance sparse matrix mathematics and interactive supercomputing has reinforced the ubiquity of power-law distributions in these observations and enabled precision measurements of the distribution parameters \cite{gadepally2018hyperscaling, reuther2018interactive, kepner2018mathematics, kepnertrillions}.  The increased accuracy of these measurements allows new underlying generative network models to be explored.

Prior modeling studies primarilyly relied on collecting data through web crawls that naturally produced connected power-law networks \cite{leland1994self, albert1999internet, barabasi1999emergence, olston2010web}.  The foundational preferential attachment (PA) random network model emerged naturally from these observations \cite{newman2001clustering, clauset2009power, barabasi2009scale, barabasi2016network, sheridan2018preferential}.  PA models rely on the connectivity of a large component to generate subsequent iterations of the model \cite{prefattach, aclpa, fanbook, ilm}. These large components, and other network cores have been studied extensively \cite{yang2020controlling,xue2017reliable, newman2001clustering, kepner2018mathematics}.  WIDE, CAIDA, and other Internet observatories provide different vantage points that can see complete streams of traffic (Figure~\ref{fig:NetworkDistribution}) that reveal significant rare leaves of connected components as well as entirely unattached links (Figure~\ref{fig:NetworkTopology}) that create deviations from standard traditional power-law models (Figure~\ref{fig:ExampleFits}). 

At a qualitative level, it is suspected that many of these leaves and unattached links are formed by bot traffic. These connections often behave in unpredictable manners, and tend to form other links only with similar (bot-like) connections. Looking beyond the noise from these bots, the predictable connections still grow and behave in a manner that satisfies the preferential attachment model.  Crucially, the internet traffic observed in pipeline is distinct from the actual long-term traffic network of the Internet. Rather, the  observed traffic is a random subnetwork of the actual network of ``who in general feels like talking with whom''. Even though over large time scales, big chunks of Internet traffic might look like preferential attachment, the observed behaviors from the data collection methods will be a random subnetwork of the underlying model. 

Our approach to understanding this subnetwork will be with a new model that considers the time iterative selection of random subnetworks of a larger data subset. This model amplifies the randomness of the unattached links and improves our understanding of the internet beyond a preferential attachment setting.  This paper will cover the introduction and explanation of the new five-parameter PALU (PA $+$ Leaves $+$ Unattached links) model, a basic analysis of the model including the degrees distribution and analysis of the variability of the model, and conclude with the results and opportunities for future work. 

\section{Network Observations}

\begin{figure}

\includegraphics[width=1.0\columnwidth]{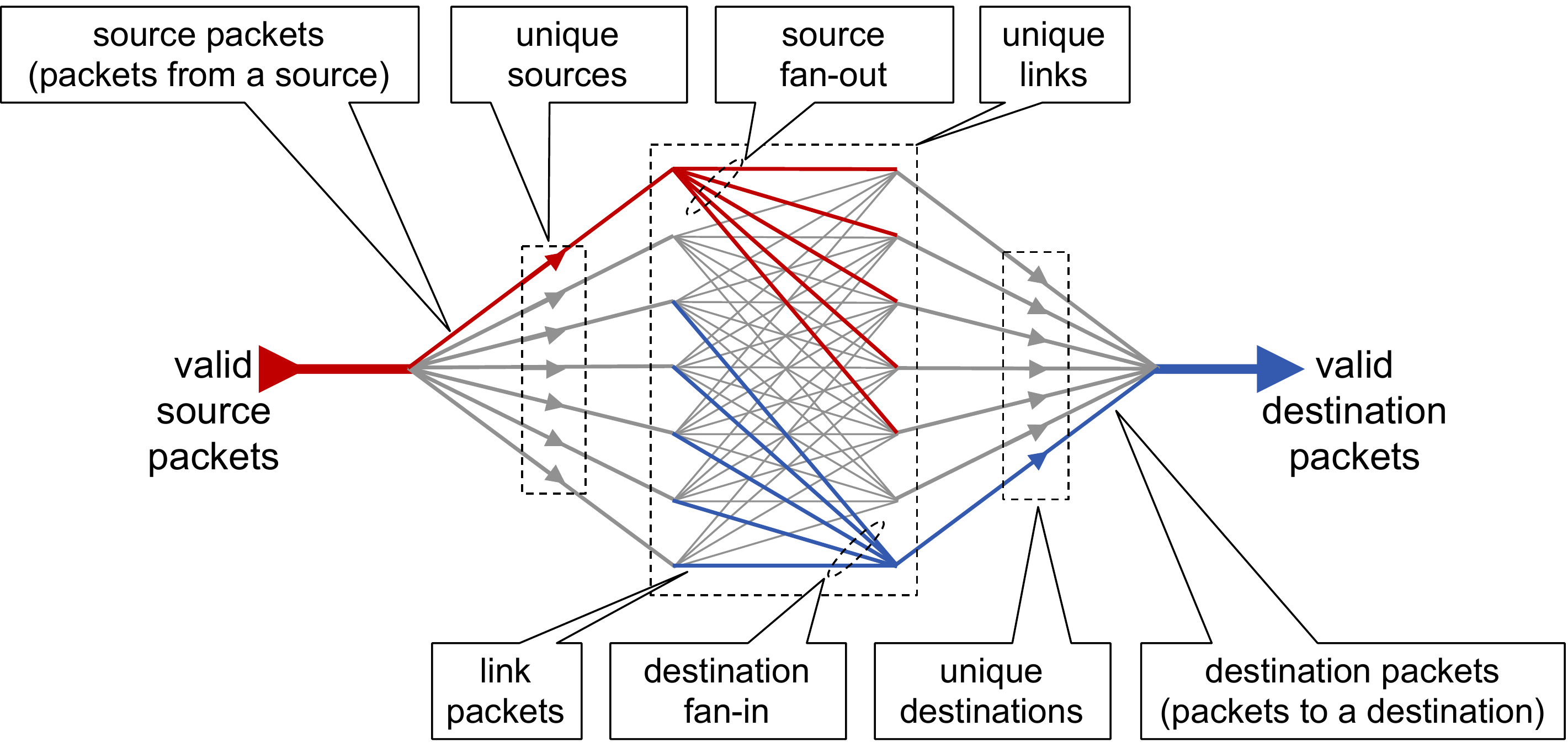}

      	\caption{{\bf Streaming network traffic quantities.} Internet traffic streams of $N_V$ valid packets are divided into a variety of quantities for analysis: source packets, source fan-out, unique source-destination pair packets (or links), destination fan-in, and destination packets \cite{kepnertrillions}}
      	\label{fig:NetworkDistribution}
\end{figure}

\begin{figure}
\centering
\includegraphics[width=0.80\columnwidth]{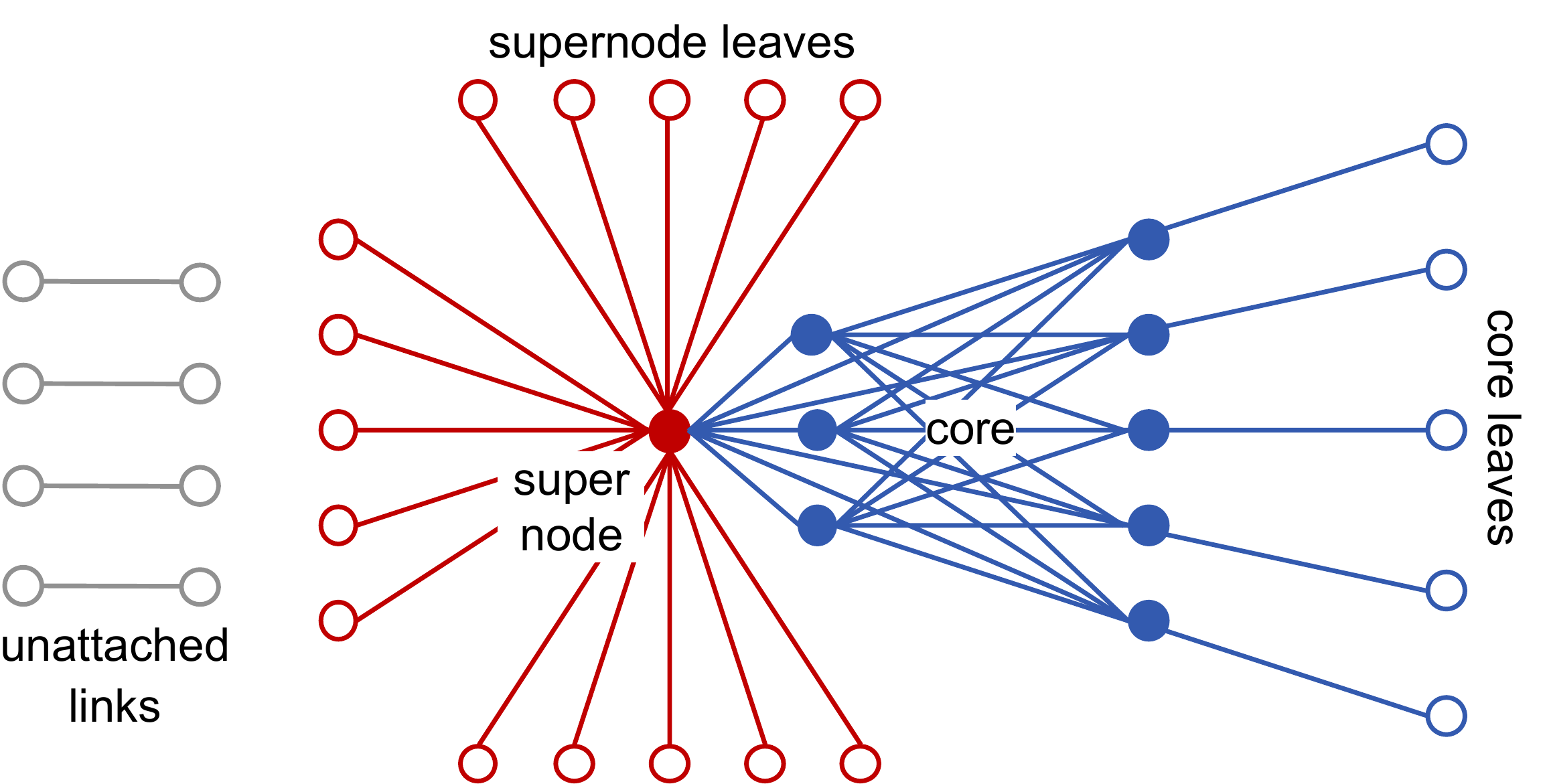}
      	\caption{{\bf Traffic network topologies.} Internet traffic forms networks consisting of a variety of topologies: unattached links, supernode leaves connected to a supernode, densely connected core(s) with corresponding core leaves \cite{kepnertrillions}}
      	\label{fig:NetworkTopology}
\end{figure}

The stochastic network structure of Internet traffic is a core property of great interest to Internet stakeholders and network scientists.  Of particular interest is the probability distribution $p(d)$ where $d$ is the degree (or count) of one of several network quantities depicted in Figure~\ref{fig:NetworkDistribution}: source packets, source fan-out, packets over a unique source-destination pair (or \emph{link}), destination fan-in, and destination packets \cite{kepnertrillions}. Amongst the earliest and most widely cited results of virtual Internet topology analysis has been the observation of the power-law relationship
$$
 p(d) \propto 1/d^\alpha
$$
with a model exponent $1 < \alpha < 3$ for large values of $d$ \cite{barabasi1999emergence, albert1999internet, leskovec2005graphs}.
In our work network topology refers to the network theoretic virtual topology of sources and destinations and not the underlying physical topology of the Internet. 
These early observations demonstrated the importance of a few supernodes in the Internet (see Figure~\ref{fig:NetworkTopology})\cite{cao2009identifying}.  Measurements of power-laws in Internet data stimulated investigations into a wide range of network phenomena in many domains and lay the foundation for the field of network science \cite{barabasi2016network}.

Classification of Internet phenomena is often based on data obtained from crawling the network from a number of starting points \cite{olston2010web}.  These webcrawls naturally sample the supernodes of the network \cite{cao2009identifying} and their resulting $p(d)$ are accurately fit at large values of $d$ by single-parameter power-law models. Characterizing a network by a single power-law exponent provides one view of Internet phenomena, but more accurate and complex models are required to understand the diverse topologies seen in streaming samples of the Internet.

At a given time $t$, $N_V$ consecutive valid packets are aggregated from the traffic into a sparse matrix ${\bf A}_t$, where ${\bf A}_t(i,j)$ is the number of valid packets between the source $i$ and destination $j$ \cite{mucha2010community}. The sum of all the entries in ${\bf A}_t$ is equal to $N_V$
$$
    \sum_{i,j} {\bf A}_t(i,j) = N_V
$$
All the network quantities depicted in Figure~\ref{fig:NetworkDistribution} can be readily computed from ${\bf A}_t$ using the formulas listed in Table~\ref{tab:Aggregates}.  An essential step for increasing the accuracy of the statistical measures of Internet traffic is using windows with the same number of valid packets $N_V$.  Using packet windows with the same number of valid packets produces aggregates that are consistent over a wide range of windows from $N_V = 100{,}000$ to $N_V = 100{,}000{,}000$. While weights are important to study, in this initial work we are studying the unweighted model. The common weights to study subsequently could be the number of packets or number of bytes sent over a link.

\begin{table}
\caption{Aggregate Network Properties}
\vspace{-0.25cm}
Formulas for computing aggregates from a sparse network image ${\bf A}_t$ at time $t$ in both summation and matrix notation. ${\bf 1}$ is a column vector of all 1's, $^{\sf T}$  is the transpose operation, and $|~|_0$ is the zero-norm that sets each nonzero value of its argument to 1\cite{karvanen2003measuring}.
\begin{center}
\begin{tabular}{p{1in}p{1in}p{0.5in}}
\hline
{\bf Aggregate} & {\bf Summation} & {\bf ~Matrix} \\
{\bf Property} & {\bf ~~Notation} & {\bf Notation} \\
\hline
Valid packets $N_V$ & $\sum_i ~ \sum_j ~ {\bf A}_t(i,j)$ & $~{\bf 1}^{\sf T} {\bf A}_t {\bf 1}$ \\
Unique links & $\sum_i ~ \sum_j |{\bf A}_t(i,j)|_0$  & ${\bf 1}^{\sf T}|{\bf A}_t|_0 {\bf 1}$ \\
Unique sources & $\sum_i |\sum_j ~ {\bf A}_t(i,j)|_0$  & ${\bf 1}^{\sf T}|{\bf A}_t {\bf 1}|_0$ \\
Unique destinations & $\sum_j |\sum_i ~ {\bf A}_t(i,j)|_0$ & $|{\bf 1}^{\sf T} {\bf A}_t|_0 {\bf 1}$ \\
\hline
\end{tabular}
\end{center}
\label{tab:Aggregates}
\end{table}%

\subsection{Logarithmic Pooling}

A network quantity $d$ computed from ${\bf A}_t$ produces a corresponding histogram denoted by $n_t(d)$, with corresponding probability
$$
    p_t(d) = n_t(d)/\sum_d n_t(d)
$$
and cumulative probability
$$
    P_t(d) = \sum_{i=1,d} p_t(d)
$$
Due to the relatively large values of $d$ observed due to a single supernode, the measured probability at large $d$ often exhibits large fluctuations. However, the cumulative probability lacks sufficient detail to see variations around specific values of $d$, so it is typical to pool the differential cumulative probability with logarithmic bins in $d$
$$
    D_t(d_i) = P_t(d_i) - P_t(d_{i-1})
$$
where $d_i = 2^i$ \cite{clauset2009power}.  All computed probability distributions use the same binary logarithmic pooling (binning) to allow for consistent statistical comparison across data sets \cite{clauset2009power, barabasi2016network}.  The corresponding mean and standard deviation of $D_t(d_i)$ over many different consecutive values of $t$ for a given data set are denoted $D(d_i)$ and $\sigma(d_i)$.

\subsection{Modified Zipf-Mandelbrot Model}

Measurements of $D(d_i)$ can reveal many properties of network traffic, such as the fraction of nodes with only one connection $D(d = 1)$ and the size of the supernode
\begin{equation}
  d_{\rm max}={\rm argmax}(D(d) > 0)
\end{equation}
Effective classification of a network with a low parameter model allows these and many other properties to be summarized and computed efficiently.  In the standard Zipf-Mandelbrot model typically used in linguistic contexts, $d$ is a ranking with $d=1$ corresponding to the most popular value \cite{mandelbrot1953informational, montemurro2001beyond, saleh2006modeling}. To accurately classify the network data using the full range of $d$, the Zipf-Mandelbrot model is modified so that $d$ is a measured network quantity instead of a rank index
$$
    p(d;\alpha,\delta) \propto 1/(d + \delta)^\alpha
$$
The inclusion of a second model offset parameter $\delta$ allows the model to accurately fit small values of $d$, in particular $d=1$, which has the highest observed probability in these streaming data. The model exponent $\alpha$ has a larger impact on the model at large values of $d$ while the model offset $\delta$ has a larger impact on the model at small values of $d$ and in particular at $d=1$.

The unnormalized modified Zipf-Mandelbrot model is denoted
$$
    \rho(d;\alpha,\delta) = \frac{1}{(d + \delta)^\alpha}
$$
with corresponding gradient
$$
    \partial_\delta \rho(d;\alpha,\delta) = \frac{-\alpha}{(d + \delta)^{\alpha+1}} = -\alpha \rho(d;\alpha+1,\delta)
$$
The normalized model probability is given by
$$
    p(d;\alpha,\delta) = \frac{\rho(d;\alpha,\delta)}{\sum_{d=1}^{d_{\rm max}} \rho(d;\alpha,\delta)}
$$
where $d_{max}$ is the largest value of the network quantity $d$.  The cumulative model probability is the sum 
$$
    P(d_i;\alpha,\delta) = \sum_{d=1}^{d_i} p(d;\alpha,\delta)
$$
The corresponding differential cumulative model probability is
$$
    D(d_i;\alpha,\delta) = P(d_i;\alpha,\delta) - P(d_{i-1};\alpha,\delta)
$$
where $d_i = 2^i$.

\begin{figure}
\includegraphics[width=1.0\columnwidth]{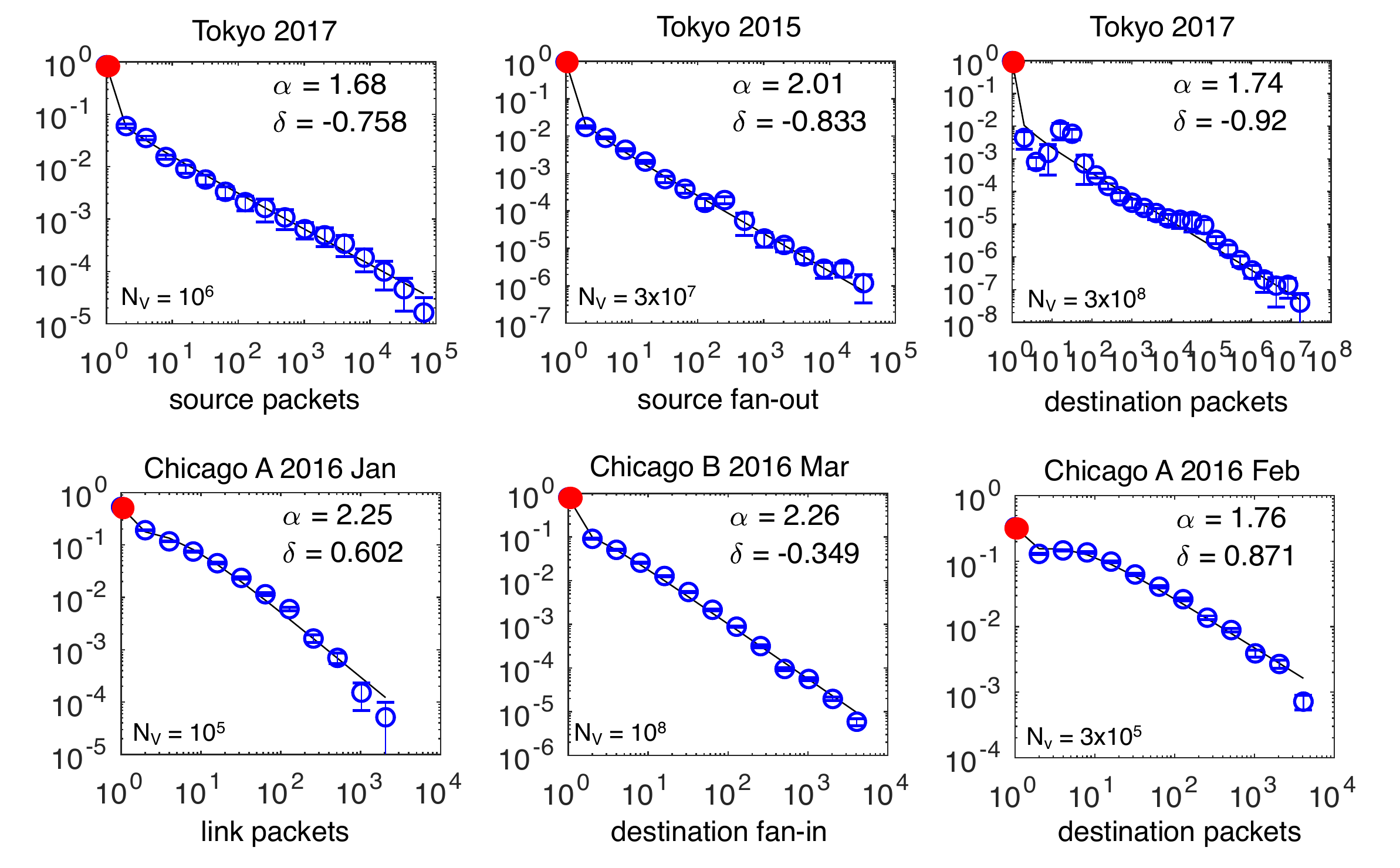}
      	\caption{{\bf Measured Distributions and Model Fits.} A selection of measured differential cumulative probabilities spanning different locations, dates, and packet windows.  Blue circles are measured data with $\pm$1-$\sigma$ error bars.  Red dot highlights the significance of leaves and unattached links.  Black lines are the best-fit modified Zipf-Mandelbrot models with parameters $\alpha$ and $\delta$ and is an excellent fit in all the examples except the upper right. }
      	\label{fig:ExampleFits}
\end{figure}

Minimizing the differences between the observed differential cumulative distributions allows accurate Zipf-Mandelbrot parameters to be selected for a given set of observations.  Figure~\ref{fig:ExampleFits} provides a representative sample of the hundreds of fits from \cite{kepnertrillions}, illustrating the effectiveness of Zipf-Mandelbrot model in describing these observations.      

\section{PA + Leaves + Unattached (PALU) Model}

The effectiveness of the empirical Zipf-Mandelbrot model drives the need to explore new underlying generative network models that can produce these more complex probability distributions.  The new model extends the PA model with explicit terms accounting for the leaves attached to the PA core as well as unattached links.

The model can be conceptualized in two parts: the \emph{underlying network}, and the \emph{observed network}. The underlying network can be considered as the ``true'' information of the traffic connections that occur and how frequent these connections are. With our current methods, the full image of this network cannot be detected, and our consideration of this underlying network is theoretical. There are three main pieces that make up this network: the \emph{core} which is constructed by preferential attachment; a set of degree 1 nodes called \emph{leaves} that are adjacent to nodes in the core; and \emph{unattached nodes} that are not connected to the core, and have very low connection within the set itself. 

Looking through a window of a certain size with respect to the number of observed connections (packets),  a random subnetwork of the underlying network is witnessed, which we call the \emph{observed network}. In this particular set of data, we observe all possible data that passes from one gateway to another until a given number of connections are observed, as described in Section II. For further details on the collection of the data see \cite{kepnertrillions}.  Conceptually, we can consider selecting this subnetwork by randomly deleting edges and nodes of the underlying network.

In reality these edge connections are  directed since internet communication is directional, however for the sake of the model we will consider this undirected. Using a directed model has a small impact on overall the degree distribution analysis \cite{barabasi1999emergence}. We will leave this discussion for later in the paper.

A webcrawl is  more likely to detect nodes in the core, but less likely to detect the leaves and the unattached nodes. The measurements taken by MAWI, CAIDA, and others through trunkline observations of trillions of connections reveal these leaves and unattached nodes included in this new model \cite{kepnertrillions}.

\subsection{Defining the Model}

The PALU model requires four parameters: the clustering of the unattached nodes, the proportions of nodes in each section of the underlying network, the preferential attachment of the core in the underlying network,  and the size of the sampled network from the underlying network.

\begin{enumerate}
    
    \item Let $\lambda \in [0,20] $ define the average degree of the unattached nodes in the underlying network
    
    \item The parameters $C,L,U$ represent the proportions of nodes in each of the core, the leaves, and the unattached nodes in the underlying network, conforming to the relationship  \[ C+L+U(1+\lambda - e^{-\lambda}) = 1 \]
    
    \item To describe the preferential attachment of the core, we require the parameter $\alpha \in [1.5,3]$ as the exponent of power-law decay of the degree distribution.
    
    \item The window size is described by the parameter $p \in [0,1]$ as the proportion of the underlying network that is being observed. As the window size increases, $p$ will get closer to 1. Specifically, this parameter is the probability that an edge in the underlying network will appear (be selected) in the observed network.
   
\end{enumerate}

The model is completely determined by five parameters, and the relationship in (1) determines the sixth ($d_{\rm max}$). Importantly, for a given network, the parameters $\lambda, C,L,U,$ and $\alpha$ should be the same regardless of the window size. As the window size increases, the only parameter that will change is $p$, as it is more likely to see more edges.

\section{Preliminary Analysis}

In the following, we reference the Riemann zeta function, which is a well-known function given by
\[
\zeta(\alpha) =\sum_{n=1} ^{\infty} \dfrac{1}{n^{\alpha}}
\]
In this paper, $\alpha$ was determined to be $1.5 \leq \alpha \leq 3$ experimentally from the log-log plot of degree distribution in \cite{kepnertrillions}.  This function is supported in MATLAB with the built-in function \texttt{zeta(x)}, and was determined here to be $1.202 \leq \zeta(\alpha) \leq 2.612$, where this will assist in the probabilistic analysis.  

Given a window size $p$, the fraction of nodes in the underlying network that we expect to see in our observed network is
\[
  V =  \dfrac{Cp^{\alpha-1}}{(\alpha -1) \zeta(\alpha)} + Lp + U(1+\lambda p - e^{-\lambda p})
\]
Our model predicts the following about the observed network, where $d$ is taken to be some integer greater than 1
\begin{eqnarray*}
\dfrac{\text{\# core nodes}}{\text{total \# nodes}} &\approx& \left(\dfrac{Cp^{\alpha-1}}{(\alpha -1) \zeta(\alpha)} \right) \dfrac{1}{V} \\ \\
\dfrac{\text{\# leaves nodes}}{{\text{total \# nodes}}} &\approx& Lp / V\\ \\
\dfrac{\text{\# unattached nodes}}{{\text{total \# nodes}}} &\approx& \dfrac{U (1 + \lambda p - e^{-\lambda p})}{V}\\ \\
\dfrac{\text{\# unattached links}}{{\text{total \# nodes}}} &\approx& \dfrac{U \cdot \lambda p \cdot e^{-\lambda p}}{V} \\ \\
\dfrac{\text{\# degree 1 nodes}}{{\text{total \# nodes}}} &\approx& \dfrac{\dfrac{Cp^{\alpha}}{\zeta(\alpha)} + Lp + U \cdot \lambda p \cdot (1+e^{-\lambda p})}{V} \\ \\
\dfrac{\text{\# degree $d$ nodes}}{{\text{total \# nodes}}} &\approx& \dfrac{\dfrac{Cp^{\alpha}}{\zeta(\alpha)} d^{-\alpha} + U e^{-\lambda p} \dfrac{(\lambda p)^d}{d!}}{V}\\
&\approx& \dfrac{\dfrac{Cp^{\alpha}}{\zeta(\alpha)} d^{-\alpha} + U e^{-\lambda p} \left(\dfrac{e\lambda p}{d} \right)^d}{V}\\ &\approx& \dfrac{\dfrac{Cp^{\alpha}}{\zeta(\alpha)} d^{-\alpha}}{V}
\end{eqnarray*}

The last two approximations are very good when $\log(d)>1$, in which case we have
\[
\log \left( \dfrac{\text{\# nodes of degree $d$}}{{\text{total \# nodes}}} \right) \approx -\alpha \log(d) + \beta
\]
where $\beta$ is a constant independent of $d$.  This provides an effective estimation for $\alpha$ via linear regression in a $\log$-$\log$ plot.

\subsection{Logarithmic Pooling}

In order to consider a selection of nodes over a logarithmically binned degree interval, simply find the cumulative sum of the corresponding estimate for all the values of $d$ within the interval.  For example with $i>3$, the number of nodes of degree between $2^{i}$ and $2^{i+1}$, we have the following summation:
\begin{eqnarray*}
\displaystyle \dfrac{1}{V}\sum_{d=2^{i}} ^{2^{i+1}} \dfrac{Cp^{\alpha}}{\zeta(\alpha)} d^{-\alpha}&=&\displaystyle  \dfrac{Cp^{\alpha}}{\zeta(\alpha) V}\sum_{d=2^{i}} ^{2^{i+1}} d^{-\alpha}\\
&\approx&\displaystyle \dfrac{Cp^{\alpha}}{\zeta(\alpha) V} \int_{2^{i}} ^{2^{i+1}} x^{-\alpha} dx\\
&=&\displaystyle \dfrac{Cp^{\alpha}}{\zeta(\alpha) V} \left(\dfrac{1-2^{1-\alpha}}{\alpha - 1} \right) \cdot \Big(2^i\Big)^{1-\alpha}.
\end{eqnarray*}
Taking logs, where $\gamma$ is some constant that does not depend on $i$, we get this is approximately
\[
   (1-\alpha) \log(2^i) + \gamma.
\]

Importantly, taking intervals of degrees for large $i$, a $\log$ plot will have the slope of the regression line as $1-\alpha$, and not $-\alpha$ as it would be in the non-interval case.  When $i$ is small (e.g., $1 \leq i \leq 3$), the approximation for the number of nodes between degree $a$ and degree $b$ with $2\leq a,b \leq 2^3$ becomes
\[
   \sum_{d=a} ^{b} \dfrac{\text{\# nodes of degree $d$}}{{\text{total \# nodes}}}
\]
and  the right-hand-side can be computed by summing up all the terms,carefully approximating with the $d!$ term and the $Lp$ term when $d=1$.  This differs from the large-$i$ estimate in that for large $i$, we can discard small terms on the right-hand-side and safely estimate the sum by an integral. In both Fig~\ref{fig:ExampleFits} and Fig~\ref{fig:palufamilies} our parameters are given for the underlying probability distribution, but we are plotting the differential cumulative distribution which will result in power law exponent one unit higher \cite{white2008estimating}.

\subsection{Simplified Degree Distributions}

The ratios for distinct degrees $d$ can be simplified using the following parameters
\begin{eqnarray}
\text{if $d = 1$,} \quad \dfrac{\text{\# degree $1$ nodes }}{{\text{total \# nodes}}} &\approx& c + l + u\label{d1}\\
\text{if $d \geq 2$,} \quad \dfrac{\text{\# degree $d$ nodes}}{{\text{total \# nodes}}} &\approx& c \cdot d^{-\alpha} + u \left(\dfrac{\Lambda}{d}\right)^d\label{smalld}\\
\text{if $d \geq 10$,} \quad \dfrac{\text{\# degree $d$ nodes}}{{\text{total \# nodes}}} &\approx& c \cdot d^{-\alpha}\label{larged}
\end{eqnarray}
where $c$, $l$, $u$, and $\Lambda$ are constants that do not depend on $d$
\begin{eqnarray*}
  c &=& C p^\alpha / \zeta(\alpha) V \\
  l &=& L p / V \\
  u &=& U \exp(-\lambda p) / V \\
  \Lambda &=& e \lambda p
\end{eqnarray*}
For $c$, $l$, and $u$, each  is proportional to the number of nodes in the core, unattached, and leaves respectively.  These three constants do depend on the parameter $p$, related to the window size.  The parameter $\Lambda$ depends on $p$ as well and is related to the clustering.  The constant $\alpha$ is the same as the above section.  All of the parameters above should be positive.

To fit these parameters, all of which are in terms of the discrete parameters, we can consider the following
\begin{itemize}
\item[(a)] One first fits \eqref{larged} to the long-term behavior of the degree distribution.  A $\log$-$\log$ plot will have a linear behavior whose slope is equal to $-\alpha$ and constant term equal to $\log(c)$, as we will see in the discussion in the following section.  This will result in fitting $c$ and $\alpha$.
\item[(b)] Subsequently \eqref{smalld} will fit small values of $d$ ($\log(d)<10$) estimating $u$ and $\Lambda$.
\item[(c)] It is then possible to solve for $l$ exactly by using \eqref{d1}.
\end{itemize}
In (b) it is possible to fit the parameters by subtracting the $c d^{-\alpha}$ term from both sides of \eqref{smalld} and then summing up both sides will give a value of roughly $u\cdot ( e^{\Lambda}-1 -\Lambda)$, which would be a more robust estimate than the point-wise estimates of \eqref{smalld}. 

After having computed $c$ and $\alpha$ from \eqref{larged}, computing \[ \displaystyle \sum_{d=2} ^{\infty} d \left[ \dfrac{\text{\# degree $d$ nodes}}{{\text{total \# nodes}}} - c d^{-\alpha} \right]\] 
should be equal to roughly $(e^{\Lambda} - 1)\Lambda u$. Thus, resulting in
\[
\dfrac{\displaystyle \sum_{d=2} ^{\infty} d \left[ \dfrac{\text{\# degree $d$ nodes}}{{\text{total \# nodes}}} - c d^{-\alpha} \right]}{\displaystyle \sum_{d=2} ^{\infty}\left[ \dfrac{\text{\# degree $d$ nodes}}{{\text{total \# nodes}}} - c d^{-\alpha} \right]} \approx \Lambda + \dfrac{\Lambda^2}{e^\Lambda -\Lambda - 1}
\]
Computing the two summations on the left-hand-side from the data can be used to to approximate $\Lambda$ by numerically solving the above.  

The advantage to this methodology is that it presumably reduces the estimate to one with substantially less variance.  Estimating the right-hand side analytically, would provide roughly accurate estimates for large $\Lambda$, but for $\Lambda \approx 0$  the estimates would become roughly $2+ \Lambda/3$, by expanding its Taylor series. After having an estimate of $\Lambda$, $c$, and $\alpha$,  \eqref{smalld} can be used to estimate $u$ by a linear regression.

A benefit to the efficacy of the model is that an arbitrary choice of $u$ and $\Lambda$ would not lead to an inaccurate computation.  In choosing these variables arbitrarily, the network would maintain the structural topology for $d=1$ since the information provided in choosing $l$ will mostly determine the structure. The model would be even more accurate for large values of $d$ by (3), and for the very few values of $d$ in between, the chosen values $u$ and $\Lambda$ could be modified until the desired results were acquired.

\section{Derivation of the estimates}

The estimates from the simplistic model follow immediately from the corresponding estimates in the refined model.  For the derivation, the aim is to count nodes rather than the ratios determined with the variables $C,L,$ and $U$. Let $N$ be the number of nodes in the underlying network. Rigorously, we let $C_N$ be the number of nodes in the core  and $L_N$ be the number of leaf nodes in the underlying network. We will carefully define $U_N$ later.  We begin the model of the observed network by beginning with the underlying network. We obtain our observed subnetwork by retaining each edge independently with probability $p$, creating an Erd\H{o}s--R\'enyi random subnetwork of the underlying network \cite{fanbook}.

 The core nodes of the underlying network are assumed to be generated according to a preferential attachment model, independent of the other parts of the underlying network.  The number of core nodes of the underlying network having degree $d$ follows a power-law distribution of the form $d^{-\alpha}/\zeta(\alpha)$.
Thus, the number of degree $d > 0$ nodes  in the observed network is well-approximated by
$$
  p d^{-\alpha}/\zeta(\alpha)
$$

Any degree $d$ node in the underlying network will have degree distributed in the observed network according to the binomial distribution
$$
  Bin(d,p) \approx d p \pm \sqrt{d (1-p) p} \approx d p
$$  

To estimate the number of core nodes with degree greater than $0$ in our observed network, we sum these estimates, which we in turn approximate as a Riemann integral, at the cost of a negligible error term.   The number of leaf nodes in the observed network has a binomial distribution with mean $p L_N$ and variance $ (1-p) p L_N$, and the random variable is approximated by its mean \cite{fanbook}.

Returning to the rigorous derivation of the unattached nodes. In this part of the model we generate $U_N$-many stars, each of which has a random number of non-central nodes, where the number of non-central  nodes is given by independent identically distributed Poisson random variables with mean $\lambda$. The Poisson random variable represents the modeling of the $U_N$ central nodes and a large number of potential leaves, each of which chooses to attach to a central node with some sufficiently low probability.  With this distribution, the total number of nodes in the unattached portion of the underlying network has a distribution precisely given by $$U_N + \sum_{j=1} ^{U_N} Po(\lambda).$$  Since the sum of independent Poisson random variables is again Poisson, this simplifies to $$U_N + Po(U_N \lambda).$$ However, of these central nodes, there will be $Bin(U_N, e^{-\lambda})$ of them which are isolated nodes.  As these cannot be seen by examining traffic between nodes,  we remove these nodes from this model. The model demonstrating a non-zero number of isolated nodes gives significant evidence to the existence of these nodes in the true picture of the internet, regardless that they cannot be observed in our current methods of data collection.

Estimating the degree sequence within our observed network amounts to the following.  If we first sample $Y \sim Po(\lambda)$ and then we take the sum of $Y$ independent $\{0,1\}$-Bernoulli random variables of mean $p$, we seek to find the resulting distribution, that is to analyze $Bin(Po(\lambda), p)$. This is well-known to be simply $Po(\lambda p)$---as seen from an elementary computation---and additionally motivated our choice of Poisson random variables.

All to say, the unattached version of our observed subnetwork is distributed as $U_N$ stars each of which have $Po(\lambda p)$ non-central leaves, from which the results follow immediately by independence and estimating the terms by their means, which is valid large values of $U_N$. 

\section{Zipf-Mandelbrot Connection}

\begin{figure}
\center{\includegraphics[width=0.55\columnwidth]{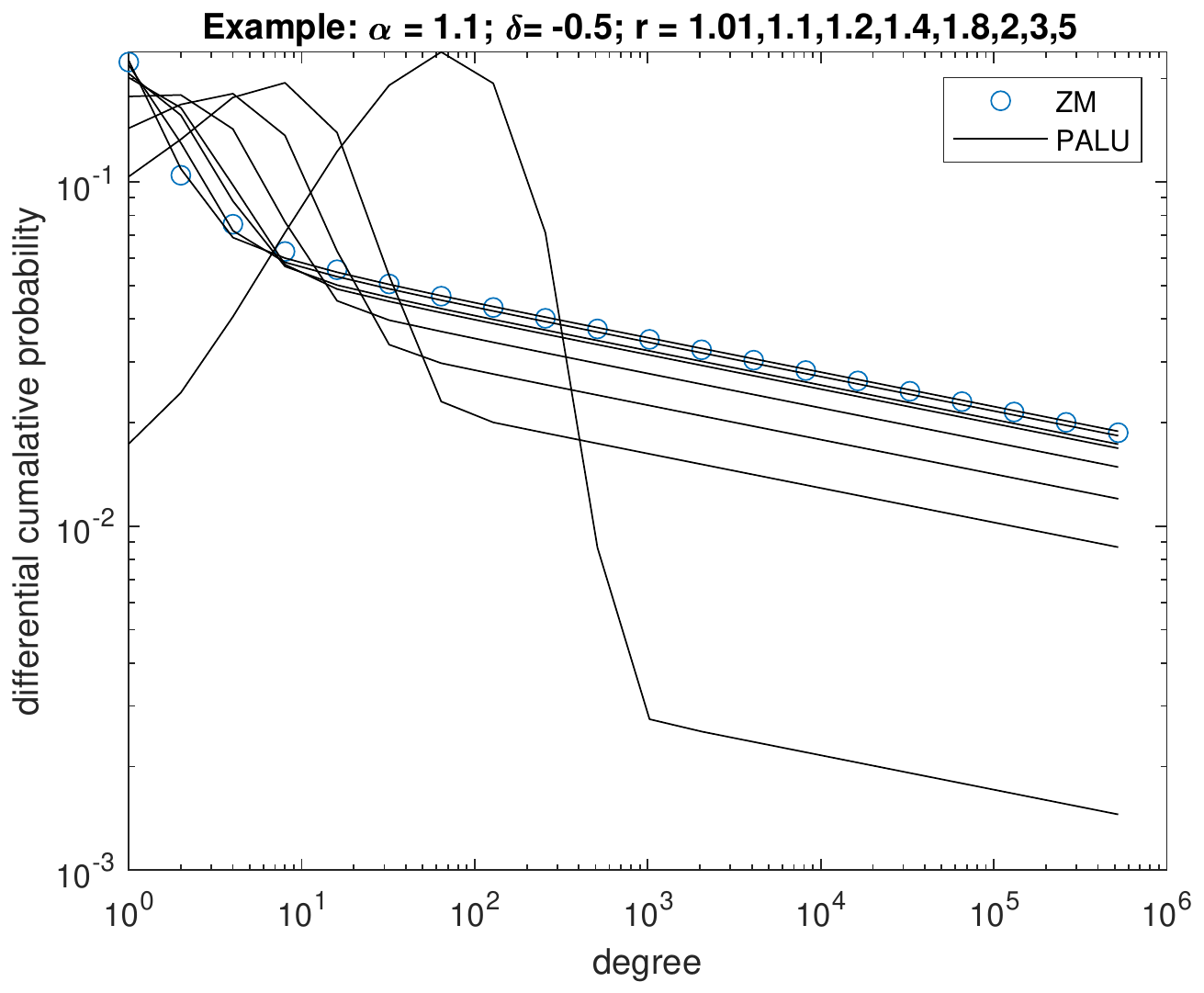}}
\center{\includegraphics[width=0.55\columnwidth]{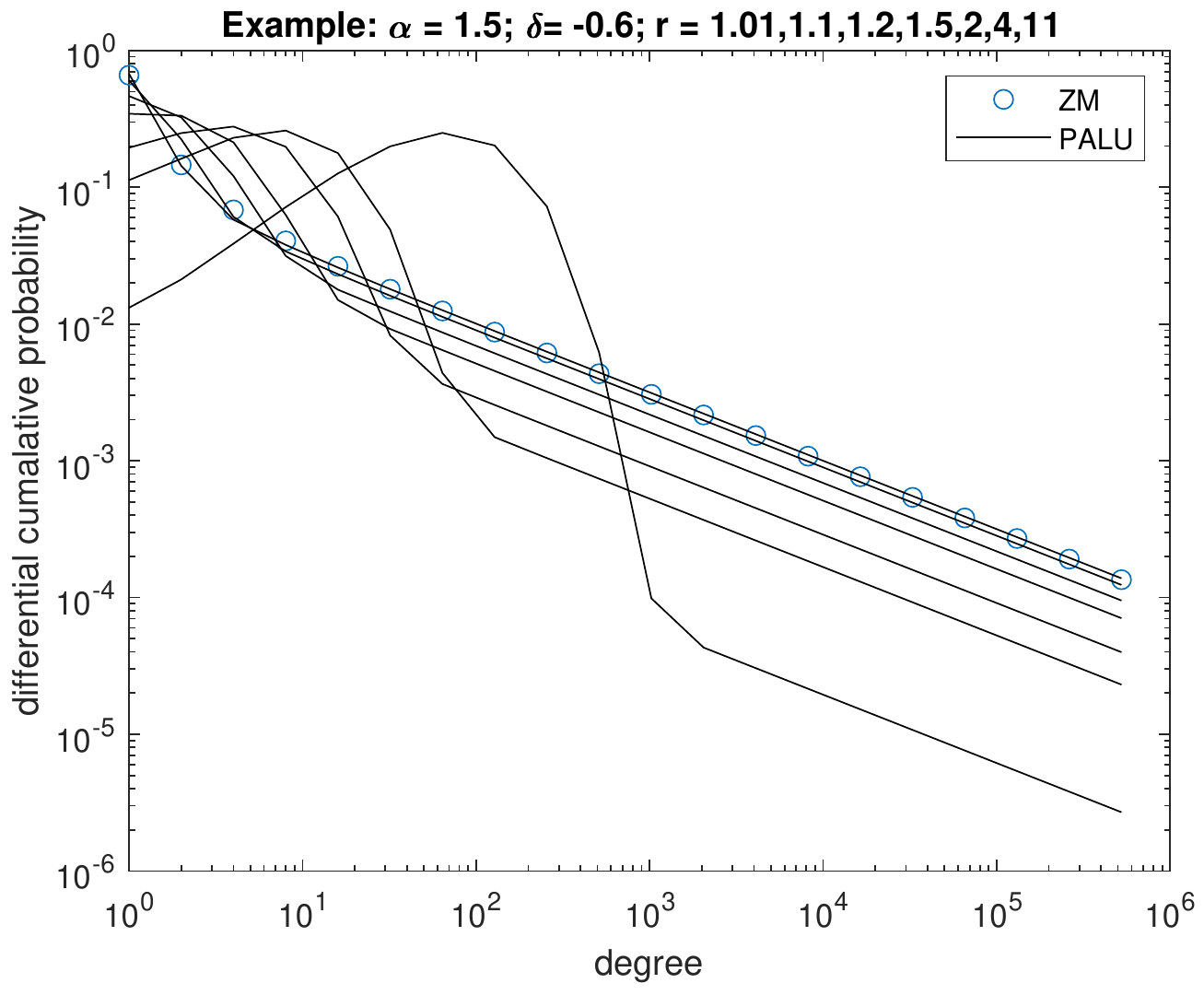}}
\center{\includegraphics[width=0.55\columnwidth]{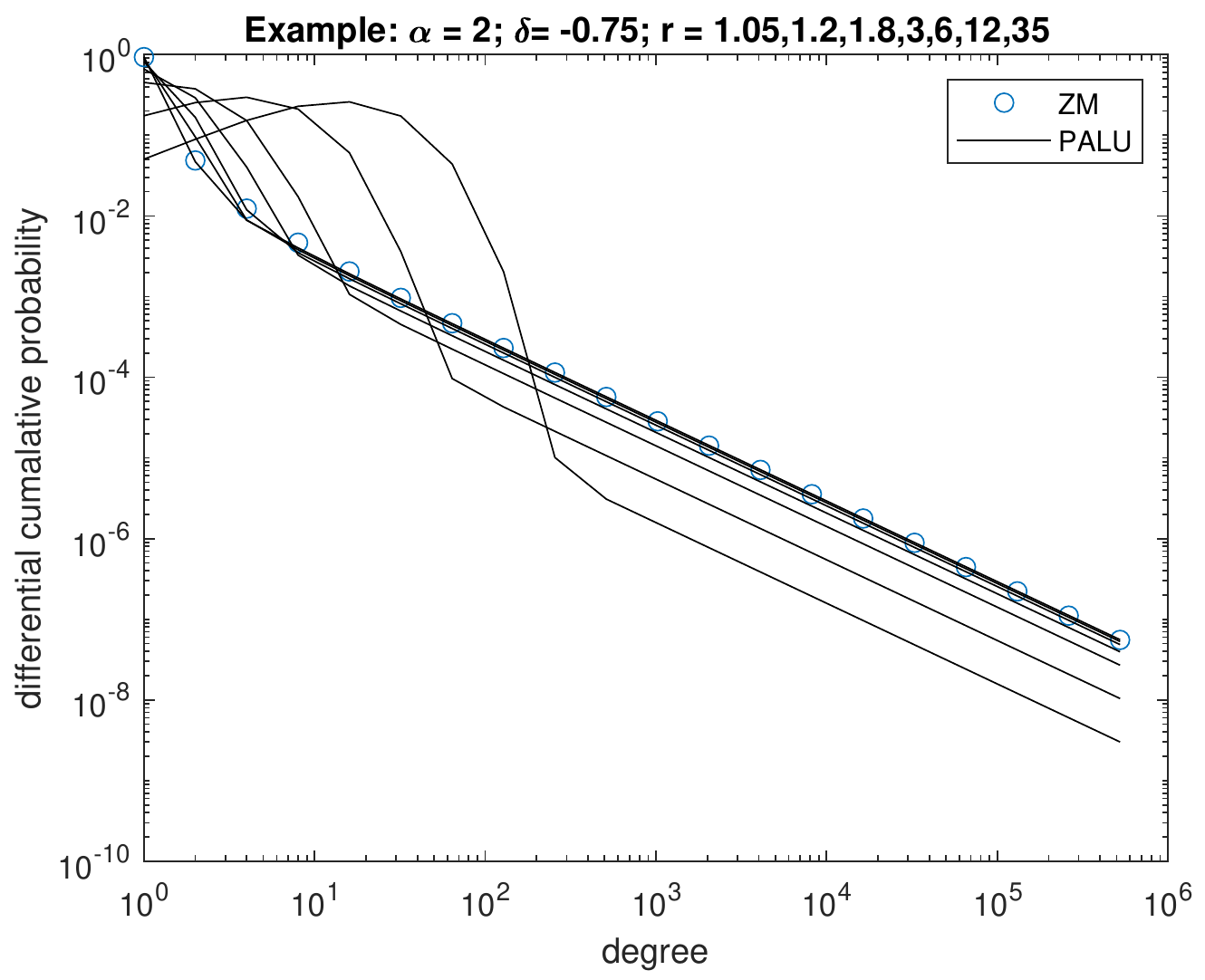}}
\center{\includegraphics[width=0.55\columnwidth]{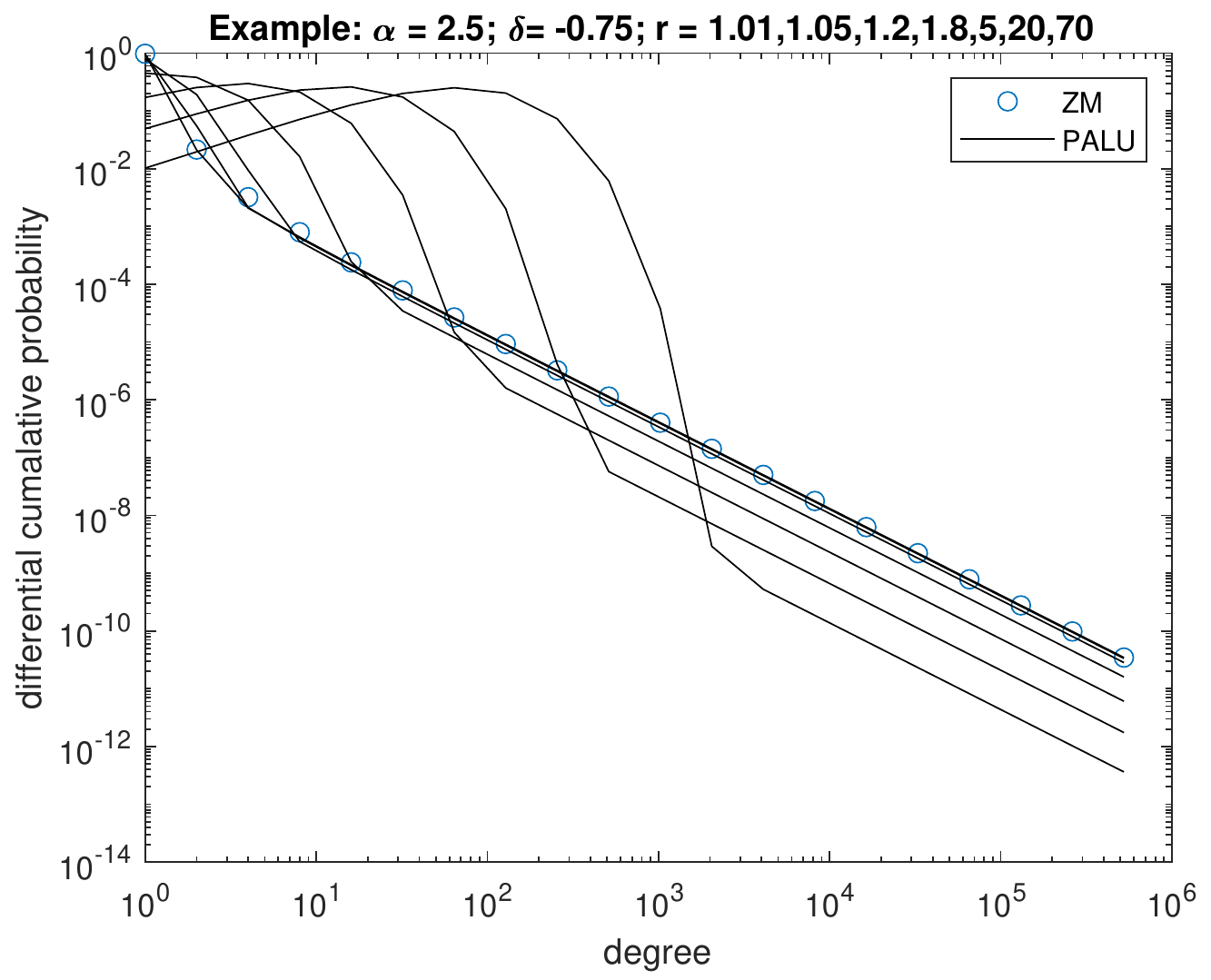}}
\center{\includegraphics[width=0.55\columnwidth]{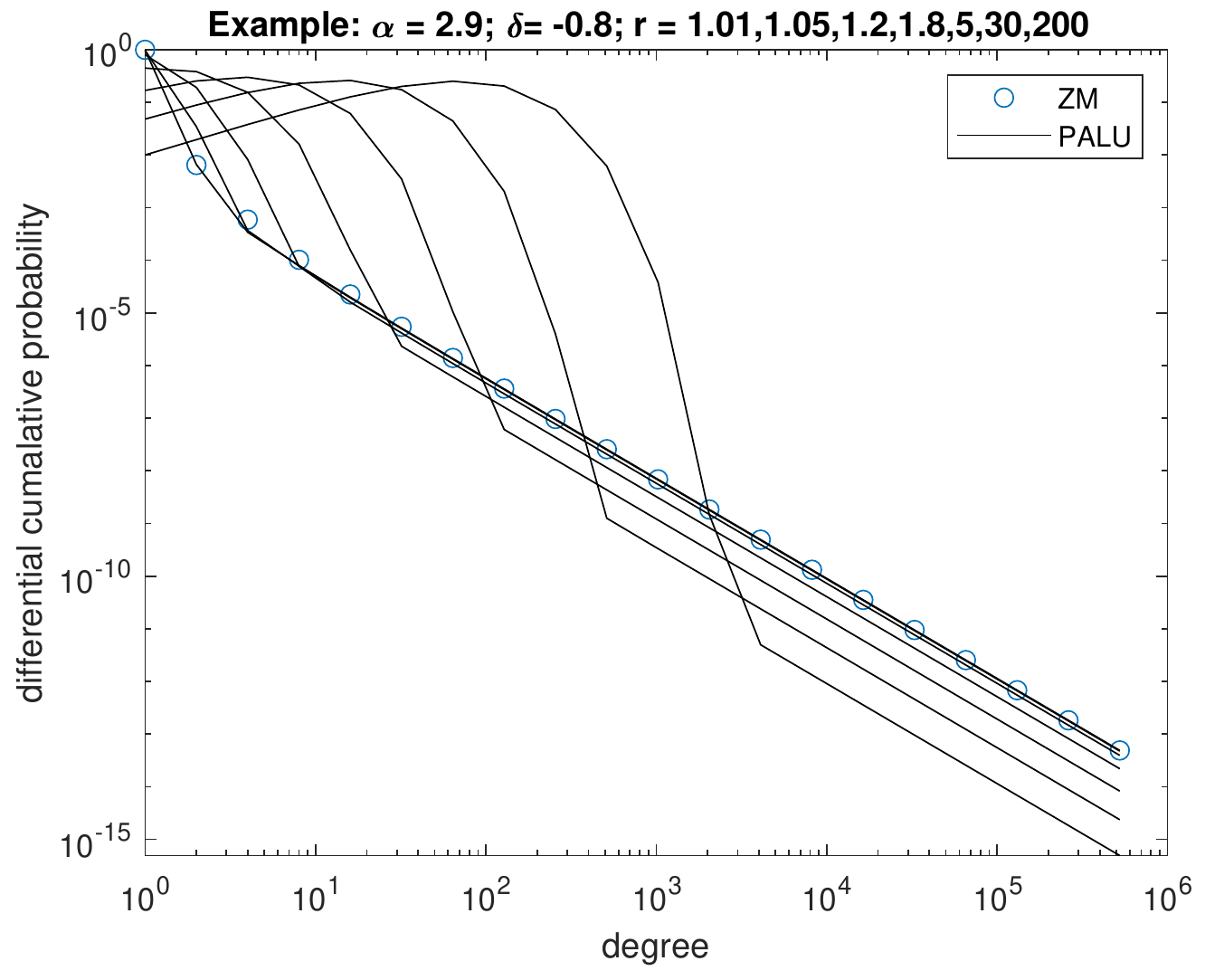}}
      	\caption{{\bf PALU model curve families.} Examples of degree distributions that are possible with the PALU model using varying parameters, and how they relate to their base Zipf-Mandelbrot (ZM) differential cumulative distributions. From the top to the bottom we vary the power law exponent $\alpha$ between 2 and 3. In each, $\delta$ is the model offset for each $\alpha$, and in each figure, we vary $r$ to create the family of curves, as described in section V with Equation~\eqref{eq:degreedisteq}. Further work will involve comparing these families with the model.}
      	\label{fig:palufamilies}
\end{figure}

Finally, a one-parameter (unnormalized) approximation of Equation~\eqref{smalld} can be used to compare with the empirical Zipf-Mandelbrot distributions.  Using the approximation
$$
  (\Lambda/d)^d \approx r^{(1-d)}
$$
where $r$ is a positive number to be fit to the data.  This approximation effectively changes the underlying distribution from Poisson to an equally valid Geometric distribution, resulting in the distribution 
$$
   c d^{-\alpha} + u r^{(1-d)}  
$$
which can be rescaled as
$$
   d^{-\alpha} + (u/c) r^{(1-d)}
$$
The above expressions can be aligned with the Zipf-Mandelbrot parameters by setting
$$
  u/c = (1+\delta)^{-\alpha} - 1
$$
resulting in
\begin{equation} \label{eq:degreedisteq}
 {\rm PALU}(d) \propto  d^{-\alpha} + r^{(1-d)}((1+\delta)^{-\alpha} - 1)
\end{equation}
In Figure~\ref{fig:palufamilies}, the ${\rm PALU}(d)$ degree distribution is shown with a selection of curve families that demonstrate the the PALU model can be made to fit a Zipf-Mandlebrot distribution for $d$.  For any given power law exponent $\alpha$ and offset parameter $\delta$, the Zipf-Mandlebrot distribution can be well-approximated by Equation \eqref{eq:degreedisteq} by varying $r$. In general, the model ${\rm PALU}(d)$ tends towards Zipf-Mandlebrot.  In addition, the ${\rm PALU}(d)$ model has the potential to explain some observations that deviate from the Zipf-Mandlebrot distribution (see Figure~\ref{fig:ExampleFits} upper right).

Connecting back to the original values of $u$ and $c$, we have
$$
  u/c = (U/C) \exp(-\lambda p) \zeta(\alpha) / p^\alpha
$$
and combining with the same $u/c$ expression in ${\rm PALU}(d)$ gives
$$
   (1+\delta)^{-\alpha} = (U/C) \exp(-\lambda p) \zeta(\alpha) p^{-\alpha} + 1
$$
Thus, the model described in Equation~\ref{eq:degreedisteq} captures the PA aspect of the model mostly
in the first term and the rest of the model is captured via $r$ and $\delta$  through the second term.

\section{Conclusions}

In this paper, we presented a modified preferential attachment model to better represent new observations of network topology observed from  streaming data. From the figures, we found the PALU model to fit the Zipf-Mandelbrot distribution very well given the right parameters. 

With the PALU model as a stepping stone, we can continue to examine the space of the network and more effective models, including comparing the data from \cite{kepnertrillions} to the model itself, extended the analysis done in Fig~\ref{fig:palufamilies}. Some opportunities for future work that we are interested in include further investigation of the PALU model such as its various applications in other big data environments and deeper study into the degree distribution and clustering coefficients. The PALU model research can also extend to the case of weighted edges where potential weights could be the number of packets or number of bytes sent along a link.  Exploration of other models are also possible, such as combining preferential attachment with the Erdos-Renyi model and determining if there is a better fitting model than the Zipf-Mandelbrot distribution.   It would be worthwhile to explore the existence and importance of isolated nodes, and analyzing aspects of the network by studying the behaviors of a random sampling or a preferential attachment graph.   Finally, extrapolating the results of the PALU model to observe and define the large clusters of small disconnected components may also be of interest.

\section*{Acknowledgments}

The authors wish to acknowledge the following individuals for their contributions and support: Bob Bond, David Clark, Nora Devlin, Alan Edelman, Jeff Gottschalk, Charles Leiserson, Mimi McClure, Sandeep Pisharody, Steve Rejto, Daniela Rus, Douglas Stetson, Allan Vanterpool, Marc Zissman, and the MIT SuperCloud team: Bill Arcand, Bill Bergeron, David Bestor, Chansup Byun, Vijay Gadepally, Michael Houle, Matthew Hubbell, Michael Jones, Anna Klein, Peter Michaleas, Lauren Milechin, Julie Mullen, Andrew Prout, Antonio Rosa, Albert Reuther, Charles Yee.

\bibliographystyle{ieeetr}
\bibliography{HybridPowerLawNetworkTraffic}

\end{document}